\documentclass[a4paper,pre,reqno,superscriptaddress,twocolumn]{revtex4}
\usepackage{graphicx,color}
\usepackage{dcolumn}
\usepackage{epsfig}  
\usepackage{soul}
\usepackage[centertags]{amsmath}
\usepackage{amsfonts}
\usepackage{euscript}
\usepackage{amssymb}
\usepackage{amsthm}
\usepackage{newlfont}
\usepackage{lipsum}
\usepackage{mathtools}
\usepackage{mathrsfs}
\usepackage{subfigure}
\usepackage{xcolor}
\usepackage{natbib}
\usepackage{hyperref}

\usepackage{xurl}

\definecolor{mypink1}{rgb}{0.858, 0.188, 0.478}

\begin{document}
 
\title[Cricket]{Individual and team performance in cricket}

\author{Onkar Sadekar}
\affiliation{Department of Network and Data Science, Central European University, Vienna, Austria.}

\author{Sandeep Chowdhary}
\affiliation{Department of Network and Data Science, Central European University, Vienna, Austria.}

\author{M. S. Santhanam}
\affiliation{Department of Physics, Indian Institute of Science Education and Research, Dr. Homi Bhabha Road, Pune 411008, India} 

\author{Federico Battiston}
\affiliation{Department of Network and Data Science, Central European University, Vienna, Austria.}

\begin{abstract}
Advancements in technology have recently allowed to collect and analyse large-scale fine-grained data about human performance, drastically changing the way we approach sports.
Here we provide the first comprehensive analysis of individual and team performance in One-Day International cricket, one of the most popular sports in the world.
We investigate temporal patterns of individual success by quantifying the location of the best performance of a player and find that they can happen at any time in their career,  surrounded by a burst of comparable top performances.
Our analysis shows that long-term performance can be predicted from early observations and that temporary exclusions of players from teams are often due to declining performances but are also associated with strong comebacks. 
By computing the duration of streaks of winning performances compared to random expectations, we demonstrate that teams win and lose matches consecutively.
We define the contributions of specialists such as openers, all-rounders, and wicket-keepers and show that a balanced performance from multiple individuals is required to ensure team success. 
Finally, we measure how transitioning to captaincy in the team improves the performance of batsmen, but not that of bowlers.
Our work emphasises how individual endeavours and team dynamics interconnect and influence collective outcomes in sports.
\end{abstract}

\maketitle

\section{Introduction}

The inception of sports, notably the Olympic Games in ancient Greece, played a pivotal role in cultural and societal bonding \cite{roche_mega-events_2000, guttmann_olympics_2002}. As societies evolved, sports mirrored changes in social structures, becoming more organised and diverse \cite{bryson_really_2008, horne_understanding_2020}. Recent digital technology advancements and enhanced data acquisition capabilities have ushered a new era of sports analytics, providing valuable insights into athlete and team performance \cite{nevill_twenty-five_2008, radicchi_who_2011, pappalardo_playerank_2019, chowdhary_quantifying_2023, zappala_paradox_2023}. In baseball, the `Moneyball' revolution popularised the strategic use of data analytics, profoundly altering team management and player evaluation \cite{lewis_moneyball_2004, brown_moneyball_2017, pappalardo_quantifying_2018}. Premier leagues in other sports such as basketball (NBA) and American football (NFL) have similarly embraced analytics to optimise player performance, team strategies, and in-game decisions, leading to stylistic shifts in play \cite{mason_putting_nodate, dawson_iron_2023, neuhaus_playing_2024}. Intellectual games like chess have also advanced with the introduction of sophisticated chess engines \cite{roring_multilevel_2007, vaci_chess_2017, chowdhary_quantifying_2023}.

One-Day International (ODI) Cricket, the world’s second most-followed sport after soccer \cite{noauthor_most_2023, noauthor_most_nodate}, enjoys widespread popularity primarily in Commonwealth countries, including India, Australia, New Zealand, the United Kingdom, South Africa, the West Indies, Sri Lanka, and Pakistan. The availability of match data, driven by amateur and professional enthusiasts, has fostered various analyses. One of the major lines of research has been predicting the match outcomes of ODI matches using a variety of techniques such as machine learning \cite{subasingha_novel_2019}, logistic regression using pre-match covariates \cite{bandulasiri_predicting_nodate, mcewan_role_2023}, and logistic regression using \textit{in-game} dynamic variables \cite{asif_-play_2016}. Other studies have tried to uncover specific patterns based on performance such as the \textit{hot-hand effect} \cite{ram_significant_2022} and ranking of the players \cite{premkumar_key_2020} or model the dynamics of the game \cite{swartz_modelling_2009}. A few studies have tried to investigate in detail the batting \cite{k_science_1905, kimber_statistical_1993} and bowling \cite{mehta_factors_1983} aspects of the game. With the advent of shorter and faster formats of the game such as T20, some attention has been devoted to investigating the effect of premier leagues on social media activity \cite{arora_indian_2014} as well as on other formats of the game \cite{nicholls_change_2023}. Going beyond the specificities of ODI, researchers have tried to examine the role of injuries in cricket \cite{j_orchard_injuries_2002, finch_measures_1999, orchard_methods_2005}. Despite this, there remains a substantial gap in the understanding of individual performances and their contributions to team success in One-day international cricket. In this work, we track players' careers, unveiling universal patterns of performance and identifying correlates of team-level success.

The progression of a player's skill level enhances their likelihood of surpassing previous performance peaks \cite{ram_significant_2022}. Conversely, as players age, their athletic prowess may diminish, potentially impeding their ability to exceed past achievements \cite{j_orchard_injuries_2002}. This contrast of skill development against physical decline poses a critical question: at which point in their careers do players deliver their best performance? We address this question in Sec.\eqref{sec:best_individual} by making use of tools and methodologies from data science and extreme value theory \cite{coles_extremes_2001, beirlant_statistics_2004} already deployed in diverse fields like science of science \cite{sinatra_quantifying_2016, liu_hot_2018,liu_general_2022}, arts \cite{fraiberger_quantifying_2018, janosov_success_2020, williams_quantifying_2019}, and sports \cite{bar-eli_twenty_2006, gilovich_hot_1985, chowdhary_quantifying_2023}.

Early identification of talent and skills can provide key advantages in many competitive settings, from firm growths \cite{golder_will_1997}, information spreading \cite{bak-coleman_combining_2022} to sports, where nurturing talent in young players can lead to higher returns \cite{baker_compromising_2018, till_challenges_2020, zappala_early_2024}. In  Sec.~\eqref{sec:peformance_prediction} we extend this inquiry to investigate the relationship between a player's initial performance and their overall career trajectory to capture whether it is possible to see hints of future performance based on early-career display.

Fluctuations in team composition frequently arise due to injuries and variations in player performance \cite{pasarakonda_how_2023}. While injuries often occur unpredictably, a decline in performance typically manifests more gradually and may not be immediately apparent. A fitting inquiry in this context is whether it is possible to discern patterns in a player's performance preceding their expulsion from the team. We study this aspect in Sec.~\eqref{sec:drop_reentry}.

The composition of an effective team encompasses skilled players under proficient leadership. Case studies across various domains, including sports, science, and business, have illustrated scenarios, where moderately weak teams achieve success under adverse conditions, guided by strategic leadership \cite{guimera_team_2005, zhang_leader_2022, hancock_good_2023, betti_dynamics_2024}. Here, we ask the reverse question, is a player's performance affected by the burden of leadership in Sec.~\eqref{sec:captain}. However, from a collective perspective, the strategic composition of a team is crucial for its effective functioning. The concept of utilising specialists -- individuals who perform highly specific roles -- extends beyond sports into various domains of human activity, including business organisations, scientific research, software development, and even hunter-gatherer societies \cite{burke_bottom_2019, salcinovic_factors_2022, wallrich_relationship_2024}. We analyse the role of three specialists -- openers, all-rounders, and wicket-keepers in Sec.~\eqref{sec:specialists}.

So far we have predominantly focused on individual contributions to team success in ODI cricket. We now broaden our perspective to analyse team success, setting aside specific individualistic factors that contribute to victory. Given that each cricket match culminates in a definitive outcome, our interest lies in discerning patterns of wins and losses for teams. We apply established metrics from the literature to quantify patterns in team success in Sec.~\eqref{sec:team_success}. Collectively, our work presents the first comprehensive analysis of individual and team performance in ODI cricket.

\section{Methods and Materials}

\subsection{Data collection}

We extracted data on 4418 One Day International (ODI) matches in men's cricket, involving 2863 unique players by web scrapping \href{http://howstat.com/cricket/home.asp}{howstat.com} (Ref.~\cite{howstat}), an open-access repository for cricket statistics, using \textit{BeautifulSoup} and \textit{urllib}, two python libraries. This dataset encompasses records of all ODI matches played from their inception in 1971 until March 2024. For each match, we extracted information including date, teams, runs scored, wickets taken, and overs played by each team, along with player names and their contributions in terms of batting and bowling. This included details like batting position, number of balls faced, number of fours and sixes hit, number of overs bowled, number of runs conceded, number of maiden overs, and number of wickets taken. Furthermore, data on the captain and wicket-keeper for each match were also collected.

Although the primary aggregation of data is at the team level for each match, the dataset is also suitable for an in-depth exploration of individual player trajectories and careers. This allows for a multifaceted analysis of performance trends, the impact of various factors on player and team success, and the evolution of the sport over more than five decades.

\subsection{Classification of players}\label{sec:player_classification}

We give a brief description of our methodology to classify players. For curious readers, we direct to Ref.~\cite{cricket_wikipedia} for the exact role of each type of player. Note that a player can be simultaneously classified into multiple categories. For e.g., a batsman can be an all-rounder, a captain, an opener, and a wicket-keeper.

\textbf{Batsman:} We consider players as batsmen who have played at least 25 matches and batted at position 7 or above in at least 50\% of their matches. We have 580  batsmen in our dataset after applying this criteria.

\textbf{Bowlers:} We classify players as bowlers who have played at least 25 matches and bowled in at least 50\% of the matches they played. We have 551 bowlers in our dataset after applying this criterion.

\textbf{All-rounders}: We consider all-rounders as players who have played at least 25 matches and are classified as both batsmen and bowlers using the above definitions. We have 117 all-rounders in our dataset.

\textbf{Captains:} In our analysis, we consider players as captains if they have captained the team in at least 15 matches. The information about captaincy is available as metadata on the website.

\textbf{Openers and Non-openers:} We characterise openers as players who have batted at first or second position at least 10 times in their career, while non-openers as players who have batted at positions three to six at least 10 times in their career.

\textbf{Wicket-keepers:} In our analysis, we consider players as wicket-keepers based on the metadata available for each match on the website

\textbf{Fielders:} All players on the bowling side who are not bowling or wicket-keeping are classified as \textit{fielder} in our analysis.

\subsection{Data normalisation}

Team scores per match have generally increased over the decades (see \textit{SM1}). This observed variation is potentially due to a confluence of factors, including modifications in-game regulations (fielding rules for example)\cite{noauthor_fifteen-over_2024}, advancements or changes in the equipment used (notably the cricket bat) \cite{sankar_evolution_2017}, and the reduction in the dimensions of the playing field 65-75 metres compared to 80-85 metres in stadiums earlier as spectators are interested in high scoring matches \cite{cricket_rules_change}. To facilitate a robust and equitable comparison of player and team performances across distinct time periods, we account for the inflation in run-scoring by implementing a normalisation procedure on all batting statistics. Specifically, we multiplied runs scored by the batsmen and the runs given by a bowler in a given year by a normalisation factor $nf = \frac{\langle Team \; runs \rangle_{all}}{\langle Team \; runs \rangle_{year}}$ such that the average team score is constant over the examined time period. Here, $\langle Team \; runs \rangle_{all}$ is the average team score across all years, while $\langle Team \; runs \rangle_{year}$ is the average team score in the given year. This procedure was originally introduced to correctly assess the impact of scientific papers published in different years \cite{radicchi_universality_2008}. Our approach ensures that the comparative analysis of players' performance from different eras is conducted in an unbiased manner that controls for background temporal trends.

\subsection{Statistical data analysis}

\subsubsection{Statistical tests}

\textbf{Linear regression} is used to estimate the linear relationship between two variables and returns a correlation coefficient. We use it in Sec.~\eqref{sec:peformance_prediction}. 

\textbf{Kolmogorov-Smirnov test} (K-S test for short) is a non-parametric test used to determine if two \textit{unpaired} sampled distributions come from the same underlying distribution \cite{hodges_significance_1958}. The null hypothesis is that the two distributions are the same and the $p$-value gives the probability that the samples in question are taken from the same distribution. We use the K-S test in Secs.~\eqref{sec:best_individual} and \eqref{sec:specialists}.

\textbf{Wilcoxon signed rank test} is a non-parametric test used to determine if the location of means of two \textit{paired} distributions is the same \cite{wilcoxon_individual_1945}. Assuming the null hypothesis that the means are the same, it returns the probability that the null hypothesis is true. We use the Wilcoxon signed rank test in Sec.~\eqref{sec:best_individual}.

\textbf{Mann-Whitney $U$ test} is a non-parametric test of the null hypothesis that the \textit{unpaired} distributions underlying the two samples are the same \cite{mann_test_1947}. This test checks the null hypothesis if one of the distributions is stochastically larger or smaller than the other. We use the Mann-Whitney $U$ test in secs.~\eqref{sec:captain}, \eqref{sec:specialists}, and \eqref{sec:team_success}.

\textbf{Welch's t-test} is used to determine if the two populations have similar means when the variances and sample sizes may or may not be equal \cite{welch_generalization_1947}. We use this test in Sec.~\eqref{sec:specialists}.

All the above tests are conducted using the \textit{SciPy} package of Python \cite{scipy_stat}.

\subsubsection{Fractional contribution}\label{sec:frac_contri}
To quantitatively and equitably assess the impact of each player, we introduce the fractional contribution $f_c$ as,
\begin{equation}
    f_c = \frac{1}{2} \times \left( \frac{\text{runs scored}}{\text{total team runs}} + \frac{\text{wickets taken}}{\text{total team wickets}} \right) \label{eq:frac_contri}
\end{equation}
, so that a player's maximum possible contribution ($f_c$) is normalised within the range of 0 to 1. It is important to note that, given this framework, specialised batsmen and bowlers have an upper limit of $f_c=0.5$ in a best-case scenario.We use this definition in Secs.~\eqref{sec:specialists} and \eqref{sec:team_success}.

\subsubsection{Effective team size}
We employ the concept of \textit{effective team size}, a well-established metric in the team analysis literature \cite{klug_understanding_2016, delice_advancing_2019} to characterise the heterogeneous contributions of individuals to the team. \textit{Effective team size} quantifies the essential number of players in a team, effectively measuring the redundancy in team composition. Mathematically, we denote the contribution of the player to the team score as $f_c$, such that $\sum_c f_c = 1$, as per Eq.~\eqref{eq:frac_contri}. The \textit{effective team size} $S_{eff}$ is defined in such a way that it equals the actual team size if contributions are evenly distributed among all players and reduces to 1 if a single player is solely responsible for the entire performance. A common approach to calculate $S_{eff}$ is to use the formula $S_{eff} = 2^{H}$, where $H = -\sum_c f_c \log_2 f_c$. Statistically, $S_{eff}$ represents the entropy of the distribution of contributions $f_i$ from team players. We use this concept in Sec.~\eqref{sec:team_success}.

\subsection{Null models}\label{sec:null-model}

\begin{figure*}[htp]
    \centering
    \includegraphics[width=1.8\columnwidth]{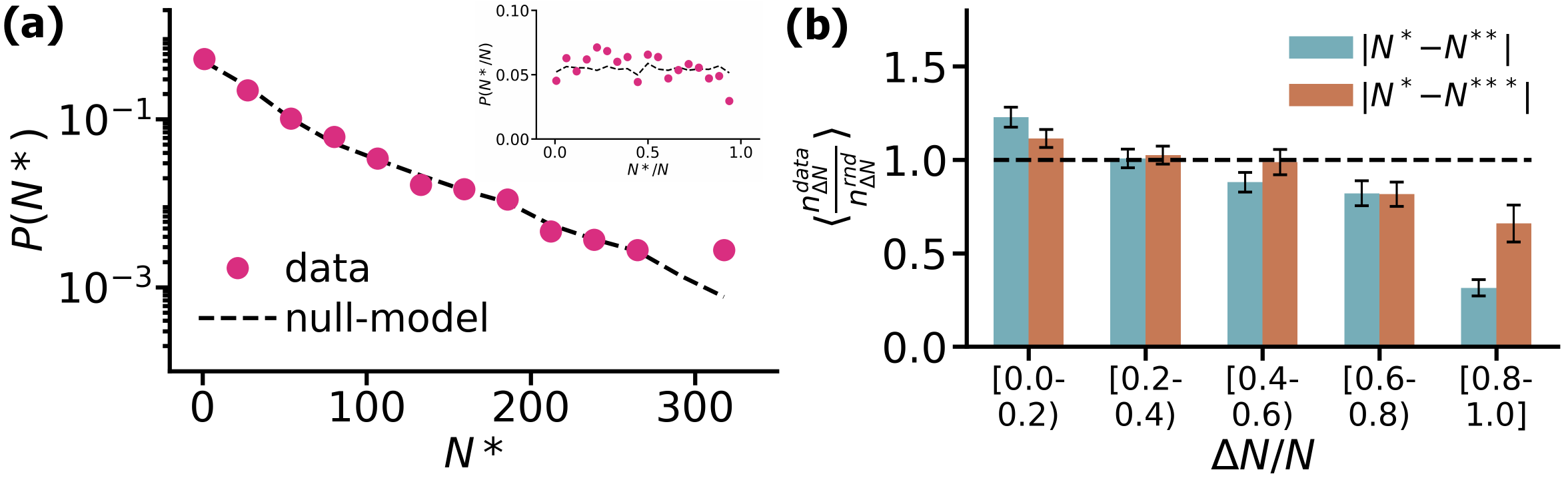}
    \caption{\textbf{Time of top performance in a player's career.} \textbf{(a)} Distribution of the match number ($N^*$) of the top performance for a player's career. The inset shows the distribution of the time of the top performance in a career $N^*$ normalised by the career length $N$. Real data (pink circles) is consistent with a randomised null model which removes temporal correlations (black dotted line) suggesting that the best performance can occur at any time within the career. \textbf{(b)} Ratio of number of players for the original data ($n^{data}_{\Delta N}$) and randomised data ($n^{rnd}_{\Delta N}$) having normalised differences ($\Delta N /N$) between best performances ($|N^*-N^{**}|$ in blue and $|N^*-N^{***}|$ in brown) in each particular bin. Values higher than 1 for small $\Delta N /N$ indicate that the best performances are clustered, highlighting the presence of \textit{hot streaks} in cricket careers.}
    \label{fig:best_individual}
\end{figure*}

In our work, we analyse characteristic patterns in the performance of both players and teams. In order to control for confounding variables, one of the usual approaches is to employ null models. Null model is a concept commonly used to capture the existence (or lack thereof) of signal from noisy data. It often involves randomizing certain aspects of the data while keeping the rest of the factors constant. Comparing the differences between the randomised and true data allows us to extract statistically significant and meaningful patterns. Here, we use two null models for detecting individual and team performance signatures.

\subsubsection{Best individual performance}

In Sec. \eqref{sec:best_individual}, we characterise the temporal patterns of top performances. In order to establish a baseline for the measures, we shuffle the timestamps of individual performances for each player 100 times keeping the actual values of performance the same. In this way, we are breaking the temporal correlation between different performances. For each shuffling, we find the time stamp for the best ($N^*$), second best ($N^{**}$), and third best ($N^{***}$) performances of a player. We compare this randomly shuffled data to the original dataset to establish the similarities and differences.

\subsubsection{Hot and cold streaks for teams}

In Sec.~\eqref{sec:team_success} we show the existence of \textit{hot} and \textit{cold} streaks for teams. We consider all the teams that have played at least 200 matches. By keeping the number of wins constant, we shuffle the timestamps of the match results $10^4$ times for each team. For each reshuffling, we compute the number of consecutive wins (hot) and losses (cold) denoted by $L_{streak}$. By taking the average across $10^4$ reshuffles, we compute $N_{rand}$ to highlight the significance of $N_{data}$ -- the number of streaks of length $l$ or higher in the actual dataset.

\section{Results}

\subsection{Temporal patterns of top performances}\label{sec:best_individual}

\begin{figure*}[thp]
    \centering
    \includegraphics[width=1.5\columnwidth]{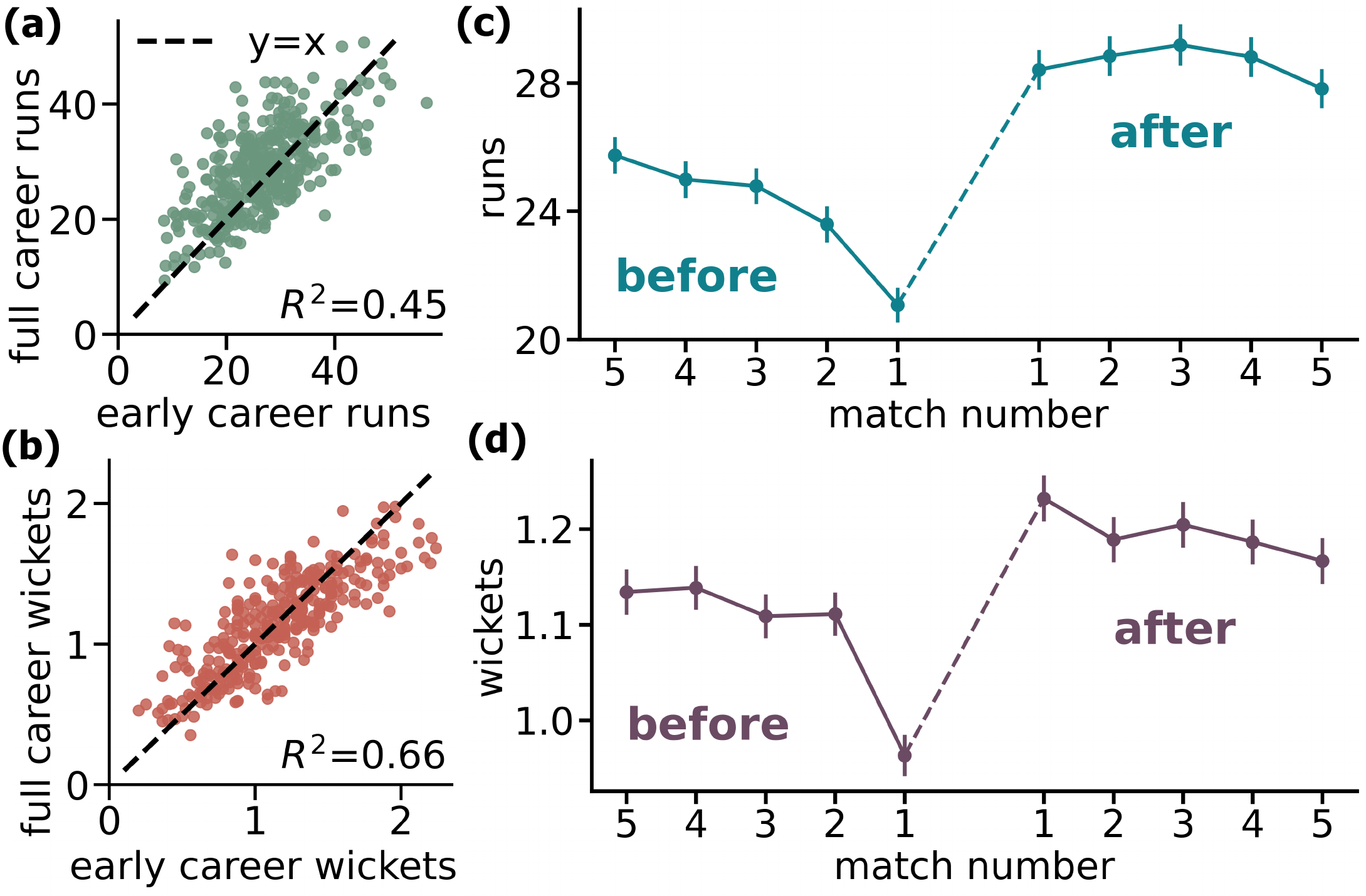}
    \caption{\textbf{Early career performance and effect of drop and re-entry} \textbf{(a), (b)} Scatter plots showing the correlation between early career and full career performance for batsmen and bowlers respectively. The $R^2$ values indicate that performances in different stages of the career are correlated. \textbf{(c), (d)} Average runs scored and average number of wickets taken by individual players (who were dropped for at least 3 matches) before they were dropped and after they were reinstated in the team. We observe a significant dip in the performance prior to the drop and a strong comeback in performance after re-entering the team.}
    \label{fig:predict_individual}
\end{figure*}

\textit{Methods:} We designate $N^*$, $N^{**}$, and $N^{***}$ as the match numbers corresponding to a player's best, second-best, and third-best performances, respectively. For batsmen, this is marked as their highest run score. For bowlers, it corresponds to the highest number of wickets taken. In instances of identical runs or wickets, the performance involving respectively fewer balls played or fewer runs conceded is considered. Additionally, to account for variations in the career lengths of the individual players, we normalise the timing of peak performances ($N^*$, $N^{**}$, and $N^{***}$) with overall career lengths ($N$).

We are interested in two questions, (1) Does the best performance ($N^*$) occur at a specific time within a player's career? and (2) Are the best performances closely related in time? To answer the first question, we calculate the probability distribution function of $N^*$ as well as $N^*/N$ and compare it with the randomised dataset as shown in Sec.~\eqref{sec:null-model}. This analysis was originally introduced in Ref.~\cite{liu_hot_2018} to test the presence of \textit{hot-streaks} in time series data. We use the K-S test to check if the distributions obtained from the original and randomised datasets differ. On the other hand, to investigate the second question, we divide the differences between the top performances $\Delta N$, i.e. $|N^* - N^{**}|$ and $|N^* - N^{***}|$, normalised by the career length ($N$) into 5 bins. Within each bin, we compute the ratios of the number of players in the data ($n^{data}_{\Delta N}$) and the number of players in the randomised data ($n^{rnd}_{\Delta N}$). We run the Wilcoxon signed rank test to determine if the values of the ratios differing from 1 are significant.

\textit{Results:}
The probability distribution function $P(N^*)$ for all players is shown in fig.~\eqref{fig:best_individual}(a). We observe that $P(N^*)$ is monotonically decreasing function, suggesting a much higher likelihood of peak performance occurring earlier rather than later in a career. However, this analysis does not account for variations in career lengths. When normalising the timing of peak performance relative to overall career length, we observe a uniform distribution, as shown in fig.~\eqref{fig:best_individual}(a) (inset). This pattern, previously dubbed as the `random impact rule' for scientific careers \cite{sinatra_quantifying_2016}, suggests that the timing of peak performance is generally unpredictable and can occur at any point in a career. The K-S test comparing the data to the null-model gives $p$-value $>0.05$ in both cases, thus giving an indication that the null hypothesis of the two distributions being statistically the same can not be rejected.

In fig.~\eqref{fig:best_individual}(b) we plot the ratios of the number of players in the bins of $\Delta N /N$ for the original and the randomised dataset. We notice that, for $\Delta N /N \in [0.0, 0.2)$ (indicating top performances closely related in time) we observe a ratio $>1$, signifying that the number of players having small gaps in their best performances is higher than expected by chance on a randomised data. In other words, they exhibit \textit{hot-streaks} -- brief periods of time accentuated by best performances of comparable magnitude. In contrast, the ratio of the number of players for $\Delta N /N \in [0.6, 1.0]$ (indicating that best performances are far away in time) is always less than one. This implies that fewer players have long gaps between best performances compared to random expectations. The Wilcoxon signed rank test indicates a significant deviation of ratios from 1 ($p$-value $<0.001$) for $\Delta N /N \in [0.0, 0.2)$ and $\Delta N /N \in [0.6, 1.0]$ for both $|N^* - N^{**}|$ and $|N^* - N^{***}|$. Our results are robust to the number of bins as seen in SM2.

\subsection{Long-term performance from early career observations}\label{sec:peformance_prediction}

We are interested in determining the relationship between a player's early career trajectory and their overall career performance. \textit{Methods:} In figures \eqref{fig:predict_individual} (a) and (b) we correlate early and overall career statistics. Our analysis considers players who have participated in a minimum of 50 matches. We focus on their mean performance during the initial 25 matches. For batsmen, we look at the average runs scored per match in this early phase, whereas for bowlers, we look at the average number of wickets taken per match within the same period. Concurrently, we examine the full-career performance of these players. 

\textit{Results:} The scatter plots in figure \eqref{fig:predict_individual} reveal a correlation between early and full career average performance metrics. We calculate the linear regression coefficients $R^2=0.45$ for (a) batsmen and $R^2=0.66$ for (b) bowlers. This finding suggests that a player's initial performance may serve as an indicator of their subsequent career performance. However, notable differences emerge when comparing batsmen and bowlers. In general, approximately 55\% of batsmen are observed to improve upon their early career averages, whereas this figure drops to about 45\% for bowlers.

\begin{figure*}[thp]
    \centering
    \includegraphics[width=1.6\columnwidth]{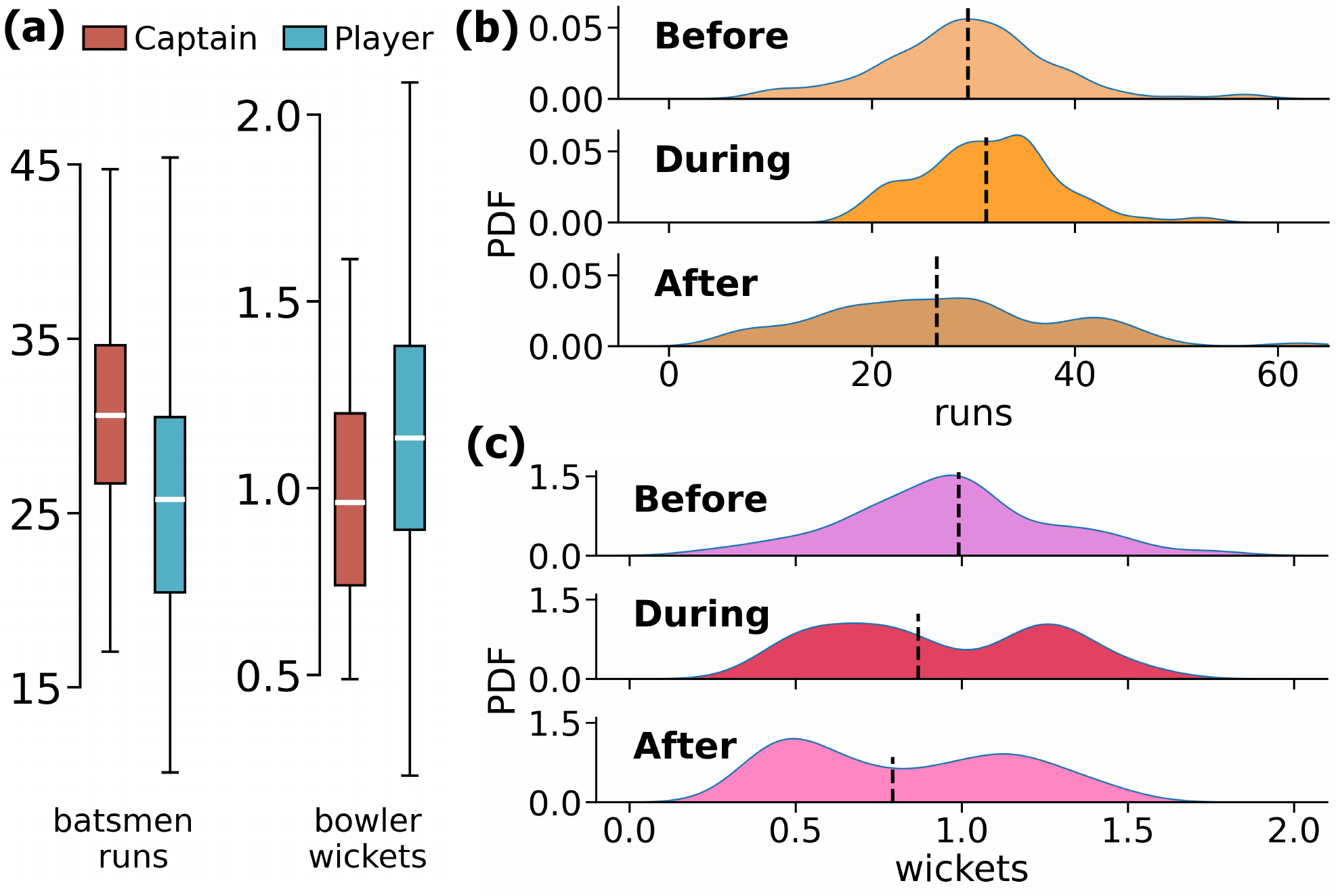}
    \caption{\textbf{Leadership affects player performance.} \textbf{(a)} Boxplot showing the average runs scored and average wickets taken by captain and non-captain players. The average trend for batsmen and bowlers is the opposite with captain-batsmen performing better than their non-captain counterparts. PDFs of performance for captains who are \textbf{(b)} batsman and \textbf{(c)} bowlers during various phases of their career. The dotted black lines denote the median of the distribution. The performance of captain-batsmen improves during captaincy, while it declines for captain-bowlers.}
    \label{fig:captain}
\end{figure*}

\subsection{Effect of drop and re-entry }\label{sec:drop_reentry}

Persistence (or lack thereof) of team composition can have a substantial effect on individual as well as team performance. \textit{Methods:} Figures \eqref{fig:predict_individual} (c) and (d) show the average performance trajectories of batsmen and bowlers, respectively, both prior to their removal from the team and subsequent to their rejoining. We consider all players who experienced a temporary exclusion from the team for a minimum of three matches before making a return.

\textit{Results:} We observe a consistent downtrend in performance during the pre-removal phase, with average runs and wickets demonstrating a monotonic decline, amounting to an approximate 19\% reduction for both batsmen and bowlers compared to 5 matches before the exclusion. Notably, the lowest performance levels are recorded in the match just before the player's exclusion. By contrast, players tend to exhibit a strong comeback performance after reinstatement to the team. In particular, the initial performance post-return exhibits a substantial elevation, with batsmen registering an approximate 36\% improvement and bowlers showing a 30\% enhancement compared to their final pre-removal performance. This enhanced performance appears to be stable in subsequent matches, consistently displaying higher performance compared to the pre-removal, suggesting that short temporal exclusion may improve performance in the longer run.

\subsection{Role of leadership}\label{sec:captain}

Proficient leadership can enhance performance in teams of skilled individuals. \textit{Methods:} We consider captains who have led their team in a minimum of 15 matches (Sec.~\eqref{sec:player_classification}. Within our dataset, we identify 172 captains, comprising 71\% batsmen and 29\% bowlers, indicating a higher propensity for batsmen to be chosen as captains -- more than twice as often as bowlers. Despite these differences, captain-bowlers remain in this position for longer than the batsmen on average (see \textit{SM5}). We first compare the career performance of captains with other players for both batsmen and bowlers. We use the Mann-Whitney $U$ test to determine if the distribution of captain performances is the same as that of the players' performance distribution. To investigate the influence of leadership, we divide the career trajectories of captains into 3 phases, before, during, and after the captainship. We are interested if the performance qualitatively changes during the three phases.

\begin{figure*}[htp]
    \centering
    \includegraphics[width=1.6\columnwidth]{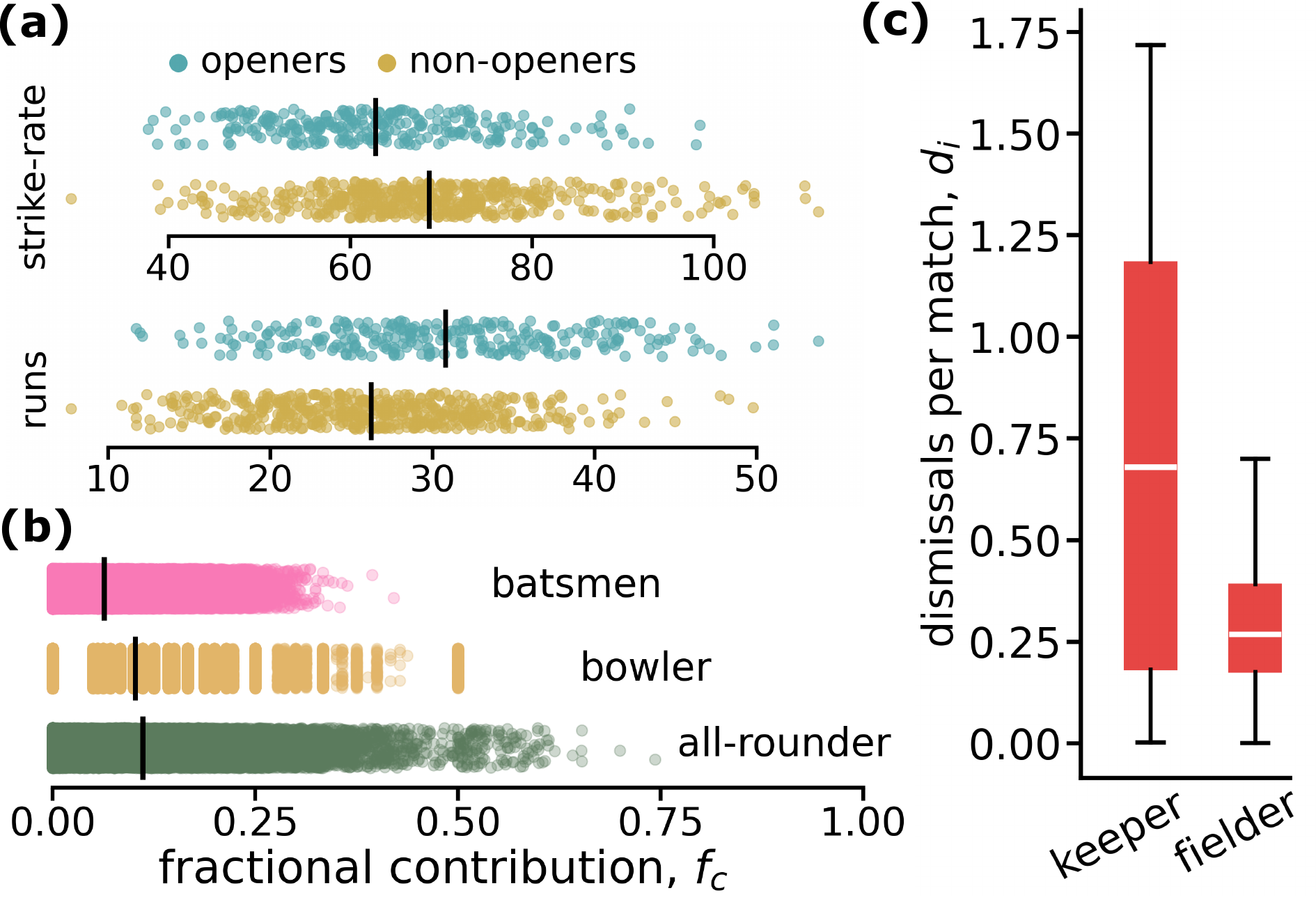}
    \caption{\textbf{Contribution of specialists.} \textbf{(a)} Runs and strike rates of opener and non-opener batsmen. The solid black lines denote the mean of the distributions. Openers score more runs at a slower pace, while non-openers play aggressively at the cost of lower contribution to team scores.\textbf{ (b)} Fractional contribution $f_c$ of players to the team scores for batsman, bowler, and all-rounder. All-rounders contribute significantly higher to the team. \textbf{(c)} Fractional dismissals $d_i$ per match by wicket-keeper and other fielders over the entire career. wicket-keepers exhibit a very high tendency to contribute to team dismissals as compared to the fielders.}
    \label{fig:specialists}
\end{figure*}

\textit{Results:} Figure \eqref{fig:captain} (a) presents a comparison of the career average performances between captains and non-captains (referred to as players). Quantitatively, the average runs scored by captain-batsmen ($31$) exceed those of players ($26$) by approximately 16\%. Conversely, captain-bowlers secure, on average, 18\% fewer wickets than non-captain-bowlers ($0.96$ versus $1.13$). We have considered only the matches where players get a chance to perform. For both batsmen and bowler, the Mann-Whitney $U$ test yields $p$-value $<0.001$. This disparity suggests differing pathways to captaincy, where batsmen may need to consistently outperform peers, while bowlers' captaincy seems less dependent on individual performance.

For both batsmen (fig \eqref{fig:captain} (b)) and bowlers (fig \eqref{fig:captain} (c)). We observe that batsmen typically enhance their performance upon assuming captaincy, whereas bowlers exhibit a decline. A closer examination of fig \eqref{fig:captain} (c) reveals a bimodal distribution during the captaincy phase for bowlers, indicating a dichotomy in skill sets among captain-bowlers. Post-captaincy, the trajectories diverge significantly for batsmen and bowlers. For batsmen, average performance markedly decreases post-captaincy, even falling below their pre-captaincy levels. The decrease is even more substantial for bowlers.

\subsection{Contribution of specialists} \label{sec:specialists}

Specialists, providing focused contributions to group endeavours are often key for team success. \textit{Methods:} Here we consider three types of specialists -- openers, all-rounders, and wicket-keepers. Batsmen are categorised as openers if they occupy the first or second position in the batting lineup, while positions 3 to 6 are considered non-openers (Sec.~\eqref{sec:player_classification}). We focus on the pace at which they accumulate these runs. The run-scoring rate is quantified as the number of runs per 100 balls faced, commonly referred to as the \textit{strike rate} in cricket terminology. This metric is frequently used to assess a player's defensive or aggressive playing style. We use the Mann-Whitney $U$ test to compare the distributions of runs (and strike rate) scored by openers and non-openers.

All-rounders are players who make significant contributions in both aspects of the game -- batting and bowling (Sec.~\eqref{sec:player_classification}). In contrast, specialised batsmen and bowlers are characterised by their significant contribution in only one of these areas. We quantify the fractional contribution for each type of player as shown in section \eqref{sec:frac_contri}. We use the K-S test to analyse the differences between the distributions of fractional contributions of batsmen and bowlers against all-rounders.

\begin{figure*}[htp]
    \centering
    \includegraphics[width=2\columnwidth]{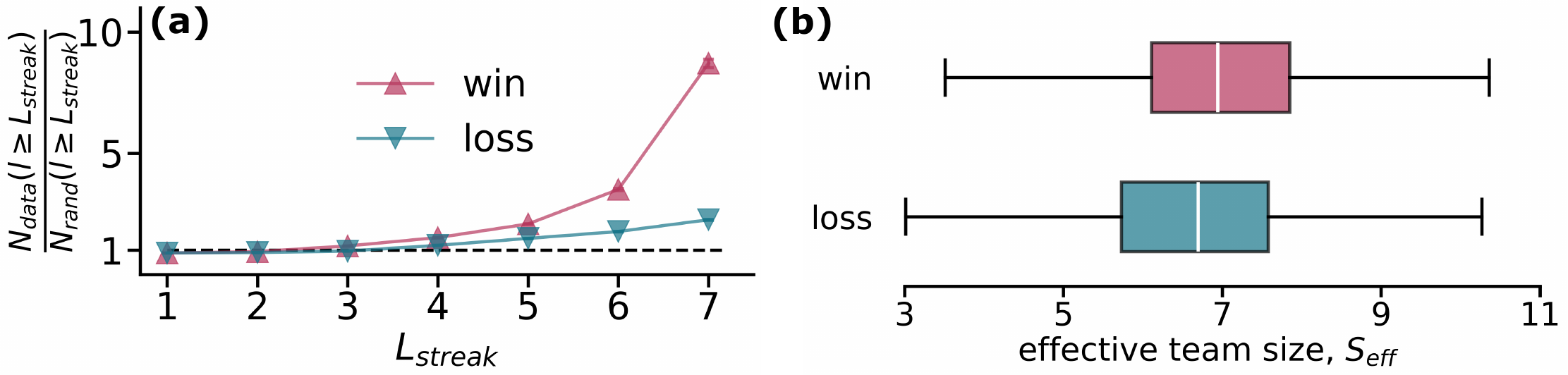}
    \caption{Occurrences of hot streaks (wins denoted by red) or cold streaks (losses denoted by blue) of a given length $L_{streak}$ compared to a null model. $N(l\geq L_{streak}$ denotes the number of times when the team had the same result for $l$ or more matches, averaged over teams. We observe that teams demonstrate both hot and cold streaks. The fluctuations in the numbers calculated as the standard error are smaller than the symbols. \textbf{(b)} Differences between \textit{effective team size} $S_{eff}$ for winning and losing teams. Winning teams have a larger $S_{eff}$ on average indicating that a larger number of players contribute significantly in winning teams than in losing teams.}
    \label{fig:team}
\end{figure*}

Among various fielding roles, the wicket-keeper is considered to hold a pivotal position. Typically, a batsman can be dismissed in three primary ways: (a) if the bowler successfully hits the stumps or the batsman obstructs the stump with their leg, (b) if a fielder catches the ball hit by the batsman before it touches the ground, or (c) if the stump is dislodged by the fielders before the completion of a run between the wickets \footnote{There are other forms of dismissals such as `retired hurt', `obstructing fielding', etc. but we do not consider them since the frequency of such dismissals is low.}. The first method relies exclusively on the bowler's skill, whereas the latter two require strategic collaboration between the bowler and fielders. Our interest lies in the last two dismissal types. Specifically, we define $d_i$ as the number of dismissals executed by a fielder of type (b) and (c) per match. We use Welch's t-test to compute the statistical significance of differences between the number of dismissals by the wicket-keepers and fielders respectively.

\textit{Results:} Figure \eqref{fig:specialists} (a) elucidates the variations in batting patterns between opening batsmen and their non-opening counterparts. Our analysis reveals that openers average a strike rate of around 63 runs per 100 balls, whereas non-openers exhibit a higher average of 69 runs per 100 balls. This suggests that non-openers score approximately 9\% more rapidly than their opening counterparts. However, the average score for openers stands at approximately 31 runs, in contrast to 26 runs for non-openers, indicating a reduction of about 17\% for batsmen coming later in the batting order. The Mann-Whitney $U$ test yields statistical significance level given by $p$-value $<0.001$.

Figure \eqref{fig:specialists} (b) presents the distribution of the contributions of all players across all matches. Among these, batsmen demonstrate the lowest mean contribution ($f_c \approx 0.06$), with bowlers exhibiting a slightly higher average contribution of $f_c \approx 0.1$. Notably, all-rounders surpass both groups with an average contribution of $f_c \approx 0.11$. Furthermore, there are instances where all-rounders singularly account for more than 50\% of the team's performance, evidenced by instances of $f_c > 0.5$. The K-S test reveals that the contributions from batsmen and bowlers are significantly different ($p$-value $< 0.001$) from those of all-rounders. We also note that non-specialists contribute significantly less than the specialists in both departments (see \textit{SM8}).

Figure \eqref{fig:specialists} (c) presents a boxplot depicting the distribution of dismissal fractions attributed to wicket-keepers and fielders. Our observations reveal that wicket-keepers play a vital role in dismissals, accounting for approximately 0.7 dismissals per match. In contrast, fielders contribute to a relatively minor number of dismissals, averaging around 0.3. The Welch's t-test gives a $p$-value $< 0.001$ implying that wicket-keepers are consistently better than fielders when the opportunity arises.

\subsection{Patterns of team success}\label{sec:team_success}

\textit{Methods:} Similar to the bursts of exceptional performance observed in individual players (as discussed in Sec.\eqref{sec:best_individual}), we assess the propensity of teams to consecutively win (or lose) a series of matches. We quantify this tendency by defining the ratio $\frac{N_{data}}{N_{rand}}$, where $N_{data}$ represents the frequency of a team winning (or losing) $l$ or more consecutive matches. In contrast, $N_{rand}$ is the analogous frequency within a randomised null model (see \textit{Methods}). Thus, the ratio $\frac{N_{data}}{N_{rand}}$ indicates the likelihood of a team experiencing a streak of at least $l$ consecutive wins or losses compared to a random outcome.

To gain more insight into the team composition, we employ the concept of effective size $S_{eff}$ as discussed in \textit{Methods}. To compute the statistical differences between the $S_{eff}$ of winning and losing teams, we use the Mann-Whitney $U$ test. Additionally, we also compute the effect size $r=\frac{U}{n_1 n_2}$, where $U$ is the test statistic and $n_1$ and $n_2$ are the sample sizes respectively.

\textit{Results:} Figure \eqref{fig:team} (a) reveals that teams are more prone to winning or losing matches in sequence than would be expected by chance. Notably, the probability of winning 7 or more consecutive matches is nine times higher than chance, while the likelihood of losing 7 or more consecutive matches is about three times the random expectation. This observation suggests that teams exhibit \textit{hot streaks} in their winning performances. Intriguingly, we also observe \textit{cold streaks}, where teams undergo multiple successive losses.

As illustrated in Figure \eqref{fig:team} (b) via a boxplot, the distributions of $S_{eff}$ for winning and losing teams differ notably. The median $S_{eff}$ for winning teams ($\approx 6.9$) is approximately 4\% greater than that for losing teams ($\approx 6.6$). This indicates that successful teams require a broader array of individual contributions. The Mann-Whitney $U$ test indicates that the differences are significant ($p$-value $<0.001$) and the effect size $r=0.56$ indicates a probability that the winning side is 56\% more likely to have higher $S_{eff}$ than the losing side.

\section{Discussion}
In the last few years, the advent of detailed sports analytics has revolutionised our understanding of human behaviour in sports \cite{lewis_moneyball_2004, brown_moneyball_2017, pappalardo_quantifying_2018}. It offers insights into physical performance, cognitive processes, and team dynamics, moving beyond traditional training methods \cite{mason_putting_nodate, dawson_iron_2023}. This shift towards a data-driven approach in sports reflects a broader societal trend in optimising human potential, combining historical practices with modern scientific methodologies.

In this study, we quantified the individual and team performance in ODI cricket, examining various aspects of the game. Due to the absence of quantifiable performance metrics such as the Elo rating system in chess, we validated the use of runs scored and wickets taken as indicators of batting and bowling performance, respectively. Our analysis revealed a significant increase in average runs scored over time. To negate this inflation in run-scoring and facilitate fair comparisons, we implemented a normalisation of all performances involving runs (see SM1).

This rescaled data corroborated the ``random impact rule" observed in other domains, indicating that a player's peak performance can occur at any point during their career \cite{sinatra_quantifying_2016, liu_hot_2018}. However, this best performance often co-occurs with other comparable exemplary performances, pointing to the existence of \textit{hot streaks} in individual cricket players.
Across a range of domains such as science \cite{liu_hot_2018, liu_general_2022}, arts \cite{williams_quantifying_2019, janosov_success_2020}, and sports \cite{ram_significant_2022, chowdhary_quantifying_2023}, individual careers display \textit{hot streaks}, where exceptional performances tend to occur in bursts which are clustered in time. Our analysis showed that, while the best performance of a cricket player's career may manifest at any point within their career, peak performances tend to occur in rapid sequence over a short period of time. This is further confirmed by the probability of performing well in consecutive matches growing exponentially as evidenced by the recurrence time distribution in \textit{SM3}.

Our exploration into the predictors of individual performance unveiled a strong correlation between early career achievements and overall career trajectory. However noteworthy differences were observed between batsmen and bowlers with batsmen improving their performance from early careers more than the bowlers. Such disparities may be attributed to the distinct elements inherent in the roles of batsmen and bowlers. Batsmen, for instance, may leverage accumulated experience and refined skills to augment run-scoring, while bowlers may depend more on innovation and strategic creativity to increase their wicket tally. This distinction potentially accounts for the observed trend of a higher proportion of batsmen surpassing their initial career performance compared to bowlers. Alternatively, our observations can also be explained if some players are simply better than the rest. Thus, their performance in various stages of their careers will correlate with their full career performance. This effect is also known as the \textit{Q-model} in the context of scientific careers \cite{sinatra_quantifying_2016}. Further analyses supporting this hypothesis are presented in \textit{SM4}. Additionally, we confirmed that players are often excluded from teams due to declining performances, yet they typically exhibit a sustained enhancement in performance following their comeback.

We investigated the relationship between player performance and captaincy. Our data suggested that batsmen who ascend to captaincy roles demonstrate higher performance levels, both overall and during their tenure as captain, in contrast to non-captain-batsmen. This pattern was not mirrored among bowlers, highlighting the differential impact of captaincy on distinct player types. These observations suggest a close correlation between individual performance and captaincy tenure for batsmen, where they often lead through exemplary performance. Conversely, bowlers may experience captaincy as a burden, reflected in their individual statistics. This could account for the observed differences in performance between captains and non-captains among batsmen and bowlers. We tested that this phenomenon is not because the captain-bowlers bowl for more time in SM6. An alternative explanation might be that captains tend to bowl at the wrong time to relieve the team pressure. However, to test this hypothesis one would need fine-grained temporal information about the progression of a match, which is not present in our data.

Focusing on the role of specialists, our findings indicated that opening batsmen typically score more runs but at a slower rate, whereas subsequent batsmen tend to adopt more aggressive strategies. Collectively, these findings illustrate a strategic balance between defensive and attacking approaches contingent on a player's batting position. Openers tend to adopt a more conservative approach, possibly due to the initial uncertainty of the pitch conditions, while subsequent batsmen often adopt a more aggressive strategy, building upon the foundational efforts of the openers. Further analysis reveals that this disparity may be partially attributed to the differing number of balls faced by players based on their batting order (see \textit{SM7}).

All-rounders enhance the team's score by applying their skills across both facets of the game. They were observed to contribute more to team performance compared to batsmen and bowlers. While bowlers generally contribute more towards the team's collective efforts than batsmen, all-rounders emerge as pivotal players, often driving the team's success with contributions that exceed 50\% of the total team scores. In the realm of fielding, wicket-keepers emerged as pivotal in effecting dismissals, thereby bolstering the impact of bowlers on team success. In summary, team success depends on the successful coordination of different types of -- often specialised -- contributions. Although most players are predominantly recognised for their batting or bowling abilities, fielding is an integral aspect of a team's overall success. Indeed, a commonly reiterated phrase in cricket states, `catches win matches', emphasising the significance of player dismissals through effective fielding. Wicket-keepers, by contributing to a substantial number of total dismissals, enhance the bowlers' efforts and, consequently, the overall team performance. These findings underscore the vital role of specialists in ODI cricket. Future research could delve into a more detailed examination of their role and their strategic impact on the dynamics of the game.

Additionally, our analysis reveals, similarly to individual performances, also teams tend to win or lose matches in clusters, thus identifying both \textit{hot} and \textit{cold} streaks during seasons. We observed that winning teams were characterised by more evenly distributed player performances, as indicated by higher \textit{effective team sizes}. A lack of such collective effort often results in teams losing matches. Moreover the data reveals a redundancy of 3 to 4 players in most teams. However, our supplementary analysis also reveals that the top performers from the winning team contribute more to their team than their top performing counterparts from the losing team (see \textit{SM9}).

Our results might be subject to certain limitations. Cricket is a multi-faceted game, and while our analysis focuses on a wide range of indicators, it still does not consider all aspects of the game such as home advantage and differences among spinners and fast bowlers. Furthermore, the volume of One Day International (ODI) matches, and the resulting dataset size is not fully extensive. Future research could benefit from incorporating data from diverse cricket formats such as Test matches, T20 games, and franchise leagues like the IPL, which might yield deeper insights into varied player performance patterns. Supplementing our analysis with data from various professional levels may reveal further patterns in players' mobility and skill levels. Additionally, our dataset focuses on end-match statistics, omitting the nuanced temporal dynamics within individual games. While the final scorecard offers substantial information, some facets of performance may only become apparent with a more granular data analysis.

All in all, our work reveals intriguing patterns of individual and team performance in One Day International cricket. We believe that our methodologies developed here could be readily applied without substantial modifications to other formats of the game such as T20 and Test cricket, as well as other team sports. Particularly suitable are teams with high specialisation such as baseball, American football, and volleyball. A comparative analysis of our results with a broader literature on sports can improve our understanding of human behaviour. We hope that our analysis motivates further research into synergistic individual efforts, in sports and more broadly in team dynamics.

\section*{Acknowledgements}
The authors thank Aanjaneya Kumar and Lorenzo Betti for the discussions which led to an improvement in the manuscript. F.B. acknowledges support from the Air Force Office of Scientific Research under award number FA8655-22-1-7025.

\bibliographystyle{apsrev4-1} 
\bibliography{references}

\section*{Code and data availability}
Code implementing the analysis presented in the paper as well as the dataset collected from the website are available at \href{https://github.com/sadekar-onkar/cricket_public}{https://github.com/sadekar-onkar/cricket\_public}.

\section*{Additional information}
Supplementary information is provided as a separate file.

\section*{Materials and correspondence}
Correspondence should be addressed to \href{sadekar_onkar@phd.ceu.edu}{sadekar\_onkar@phd.ceu.edu} or \href{battistonf@ceu.edu}{battistonf@ceu.edu}.

\cleardoublepage
\newpage
\onecolumngrid

\section*{\large{Supplemental material for ``Individual and team performance in cricket"}}

\setcounter{figure}{0}
\setcounter{table}{0}
\setcounter{equation}{0}
\setcounter{section}{0}

\renewcommand{\thefigure}{S\arabic{figure}}

\renewcommand{\theequation}{SM.\arabic{equation}}
\renewcommand{\thesection}{SM\arabic{section}}

\setcounter{secnumdepth}{4}

\section{Inflation of runs over time}

\begin{figure}[!ht]
    \centering
\includegraphics[width=0.45\columnwidth]{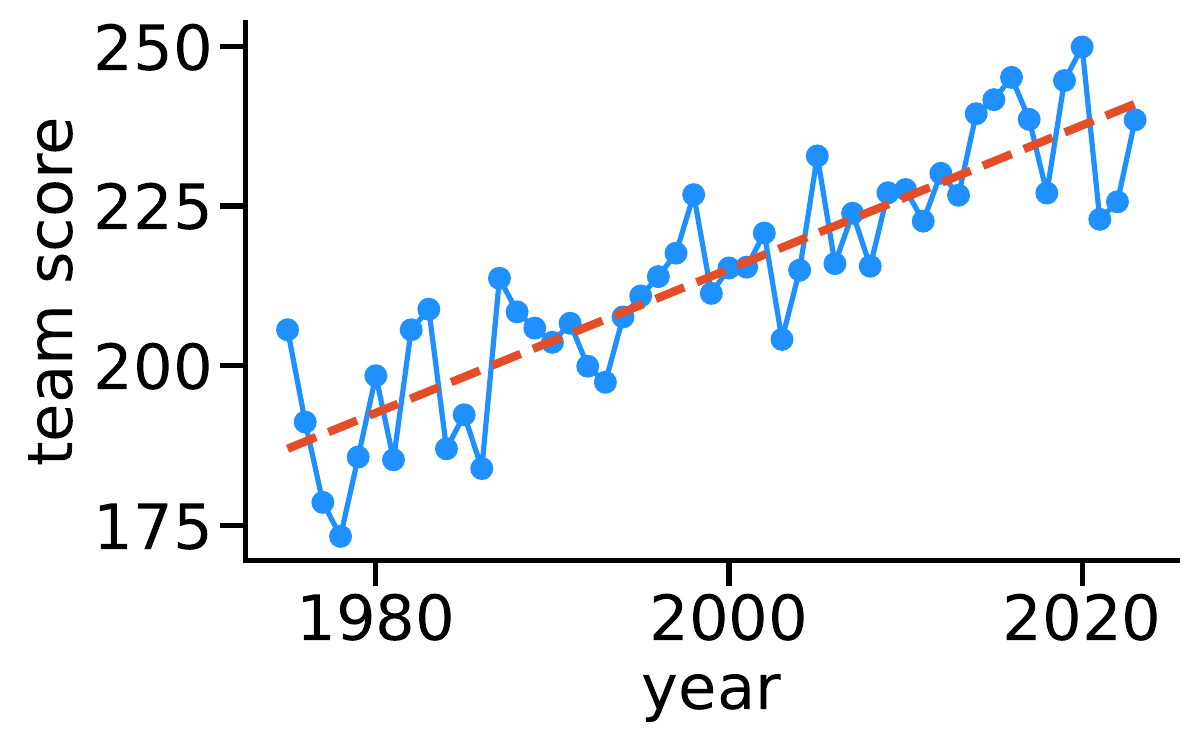}
    \caption{Inflation of performance across decades. We observe a 40\% increase in the total team score from the 1980s to the 2020s. To ensure a fair comparison for players from different eras we normalise all the scores such that the average team score over the years is constant.}
    \label{fig:run_inflation}
\end{figure}

\section{Presence of individual hot streaks}
\begin{figure}[h]
    \centering
    \includegraphics[width=0.7\columnwidth]{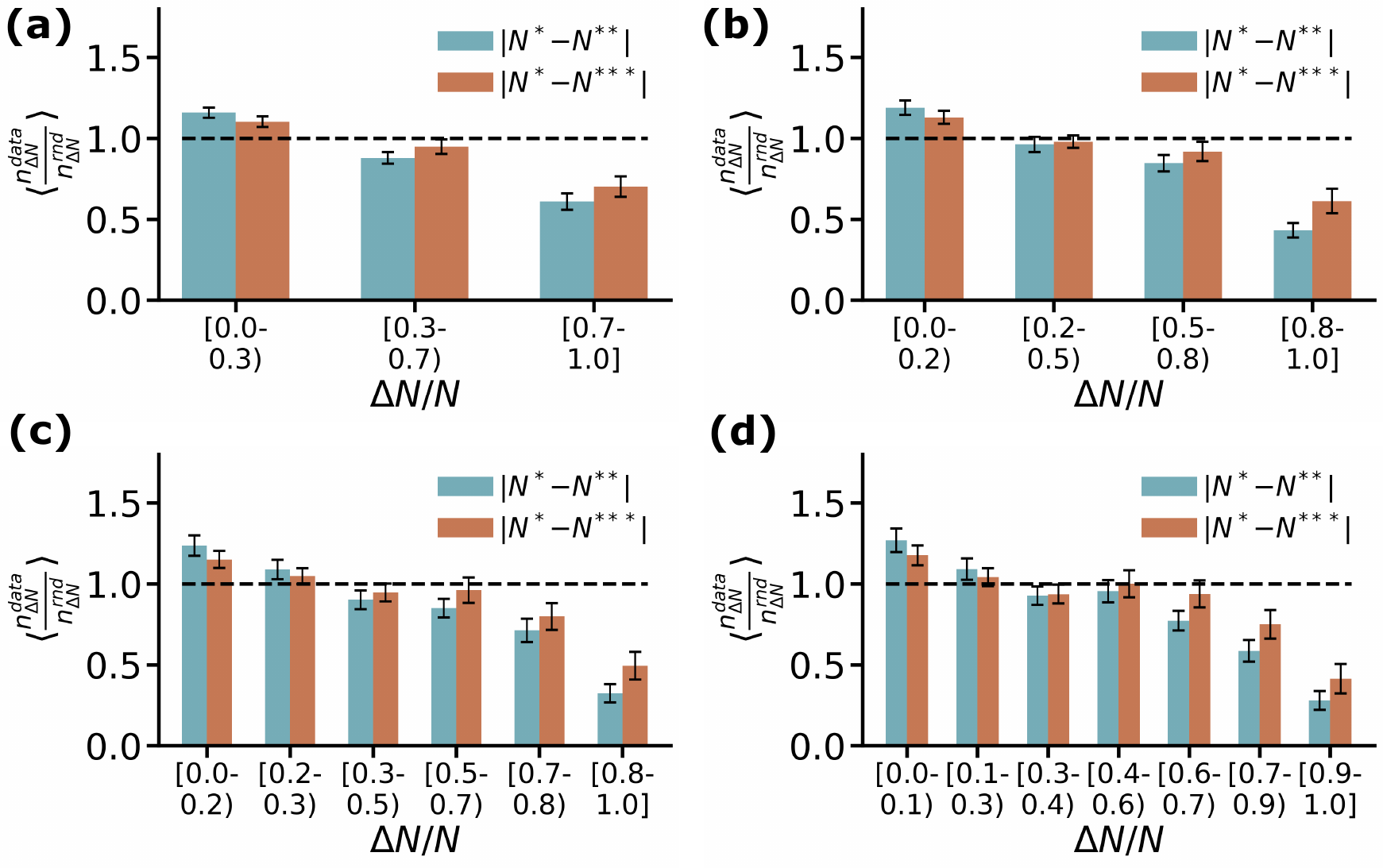}
    \caption{Presence of hot streaks for players for 4 different values for the number of bins of $\Delta N/N$. The ratio of original data ($n^{data}_{\Delta N} / n^{data}_{ N}$) and randomised data ($n^{rnd}_{\Delta N} / n^{rnd}_{ N}$) is always greater than 1 for the smallest bins, while it is always smaller than 1 for the largest bins indicating that the top performances tend to be more clustered than expected at random. All the results are statistically significant based on the Mann-Whitney U test.}
    \label{fig:indi_hot_streak}
\end{figure}

\newpage
\section{Distribution of inter-event match difference}
\begin{figure}[!ht]
    \centering
    \includegraphics[width=0.8\columnwidth]{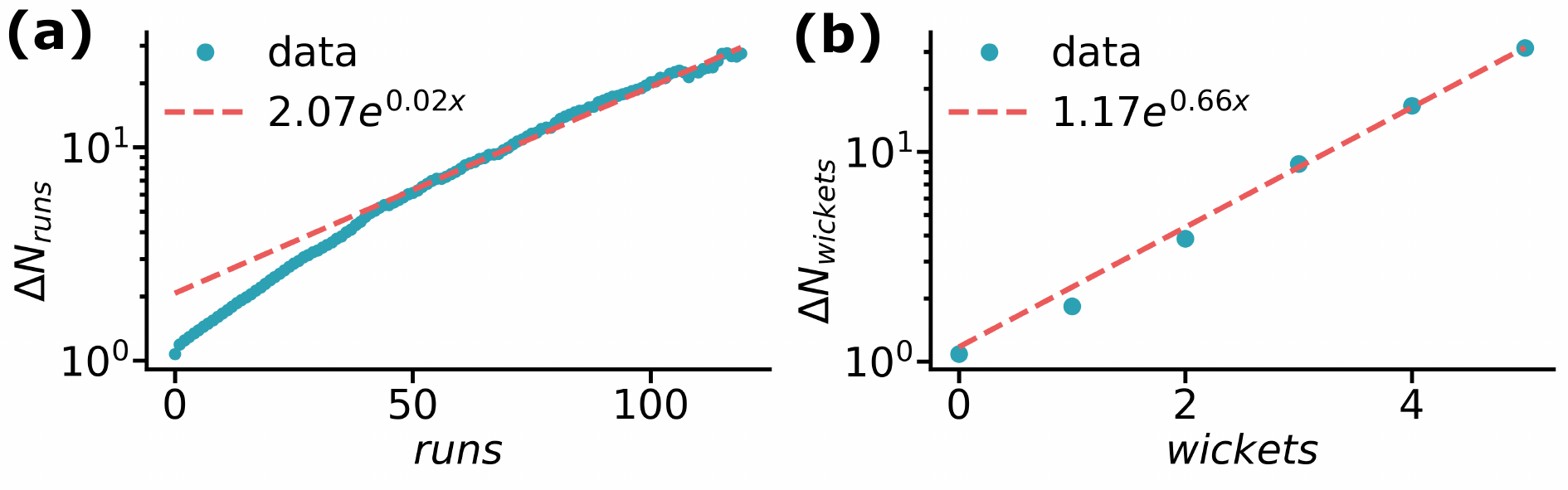}
     \caption{Distribution of recurrence times for \textbf{(a)} batsmen and \textbf{(b)} bowlers. $\Delta N_{x}$ denotes the distribution of the number of matches between consecutive instances of a player showcasing a performance of $x$ or higher for various values of $x$. For batsmen, $x$ denotes the runs scored, while for bowlers, $x$ indicates the number of wickets taken. An exponential fit reproduces this behaviour accurately, implying that these extreme events are rare (e.g. a batsman takes on average 15 matches to score 100 or more runs after scoring 100 runs).}
    \label{fig:recurrence_dist}
\end{figure}

\section{Correlators of career performance}
\begin{figure}[!h]
    \centering
    \includegraphics[width=0.8\columnwidth]{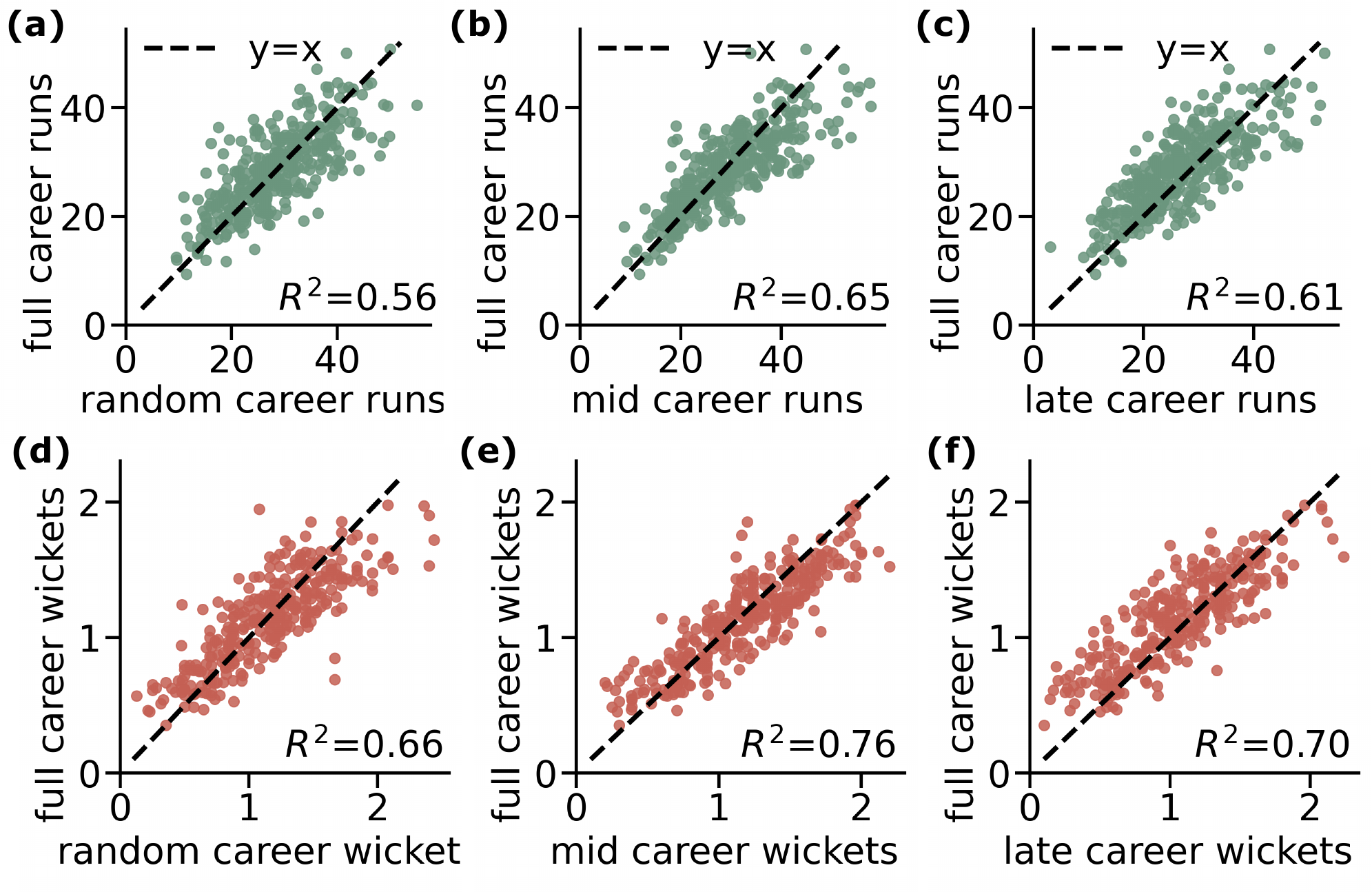}
    \caption{Correlation of mean performance over different time periods to the full career performance for (top row) batsmen and (bottom row) bowlers. We notice that \textbf{(a),(d)} random, \textbf{(b), (e)} mid, and \textbf{(c),(f)} late career performances are correlated ($R^2 \sim 0.65$) to the full career performances giving evidence that intrinsic talent influences player performance.}
    \label{fig:q_model}
\end{figure}

\newpage
\section{Length of leadership}
\begin{figure}[!h]
    \centering
    \includegraphics[width=0.75\columnwidth]{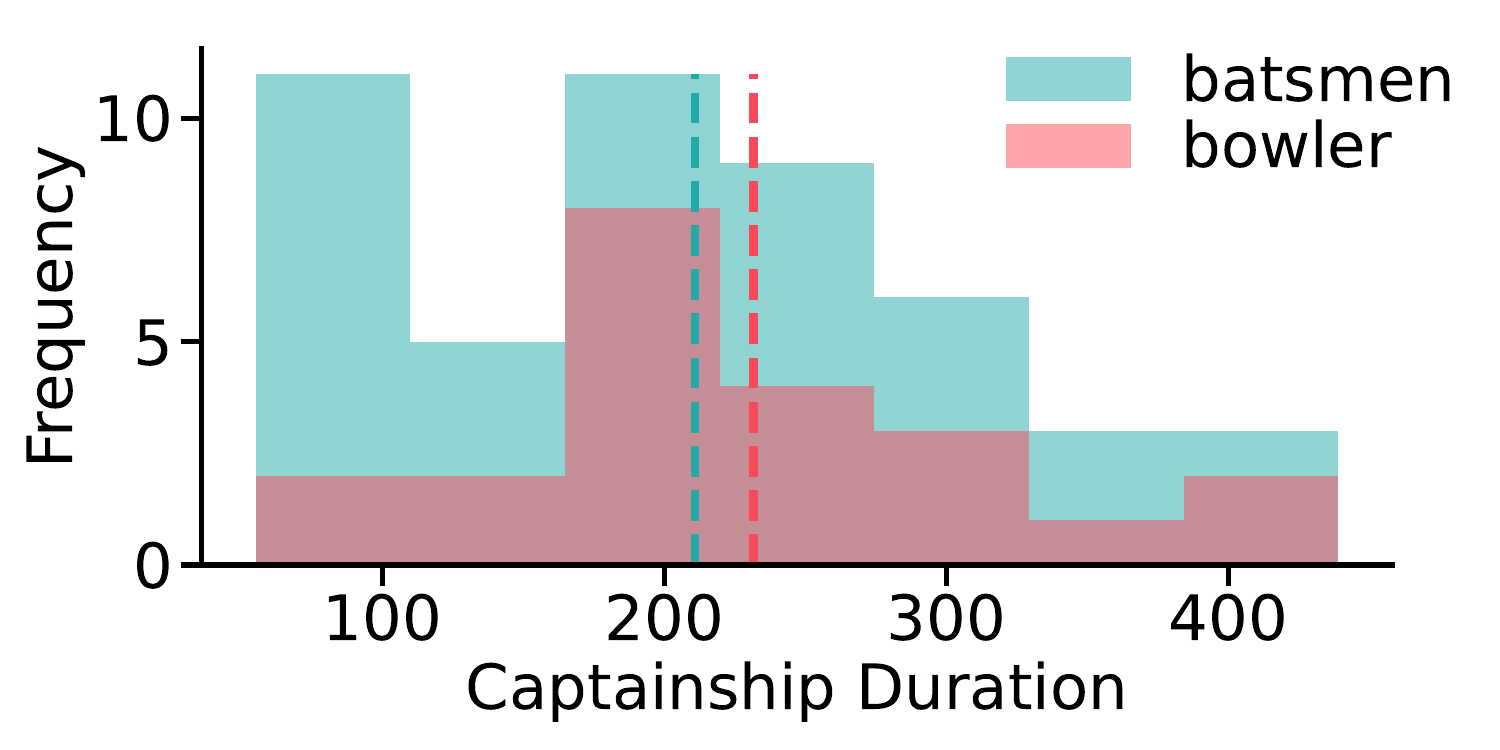}
    \caption{Distribution of leadership duration for captains who have led their team for at least 15 matches for batsmen and bowler captains. On average the batsmen lead their team in 211 matches, while the bowlers lead their team in approximately 232 matches before stepping down.}
    \label{fig:captain_time}
\end{figure}

\section{Bowler-captains' burden}
\begin{figure}[!h]
    \centering
    \includegraphics[width=0.7\columnwidth]{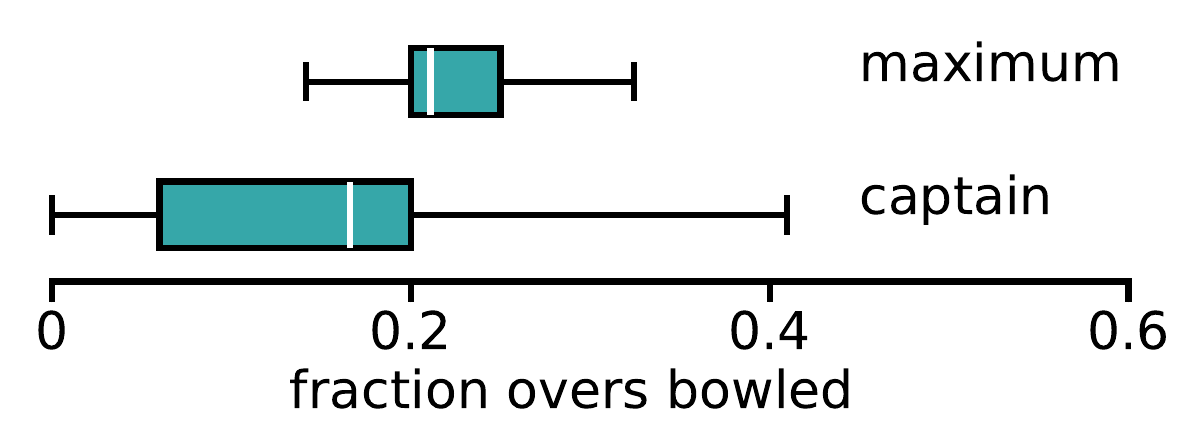}
    \caption{Boxplot showing the distribution of fraction of overs bowled by the captain (bottom) and the bowler who bowls the maximum number of overs in a given match (top). Only 31\% of times the captain bowled the maximum number of overs. The average fraction bowled by the most active bowler is 0.23, while the captain bowler bowls on average 0.14 fraction of total overs. The Mann-Whitney U test to distinguish the two distributions yields a p value $< 0.001$.}
    \label{fig:indi_hot_streak}
\end{figure}

\newpage
\section{Effect of batting position on runs scored and balls faced.}
\begin{figure}[!h]
    \centering
    \includegraphics[width=0.7\columnwidth]{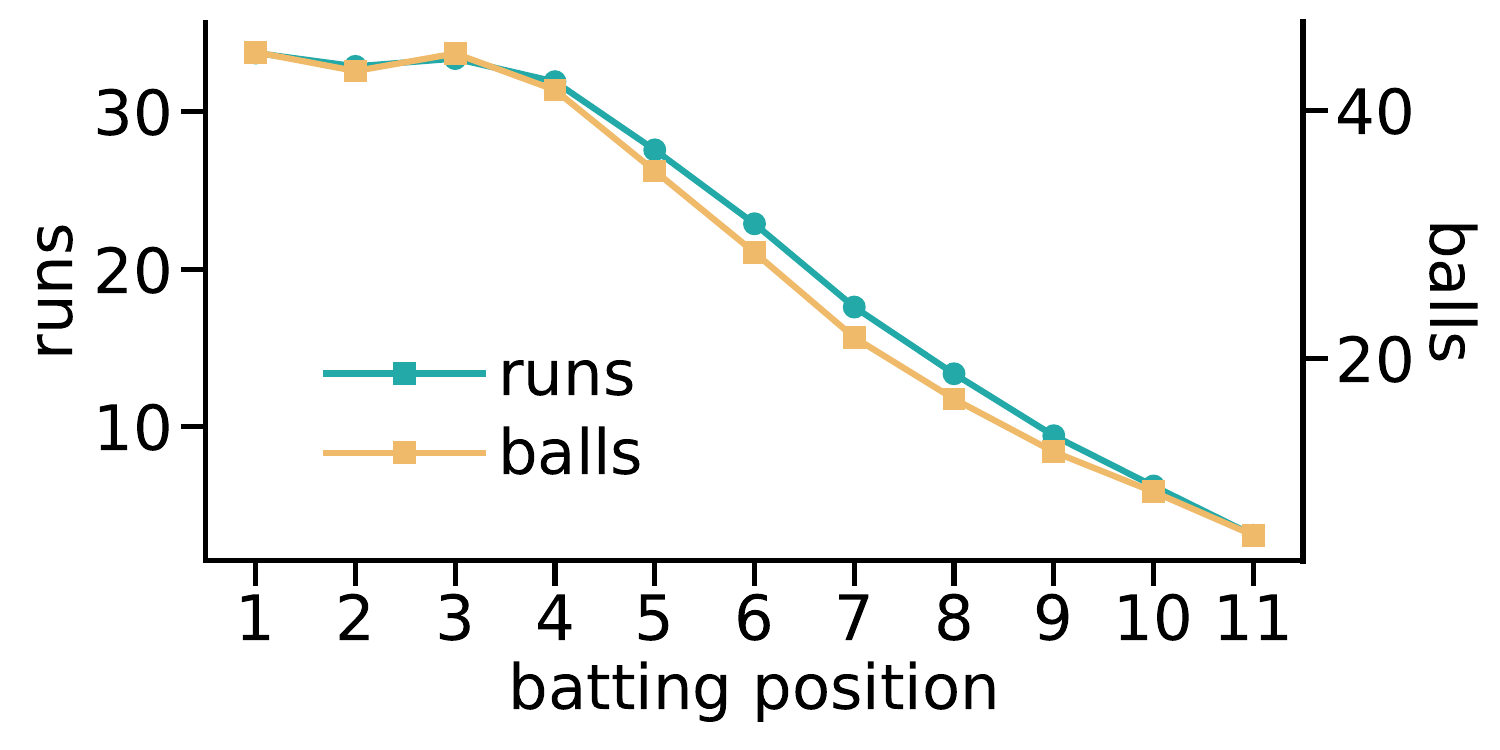}
    \caption{Number of balls faced and number of runs scored for each batting position. We notice that batsmen at positions 1,2, and 3 face more balls and score more runs than players coming in to bat at latter positions.}
    \label{fig:bat_pos}
\end{figure}

\section{Contribution of non-specialists to team}
\begin{figure}[!h]
    \centering
    \includegraphics[width=0.6\columnwidth]{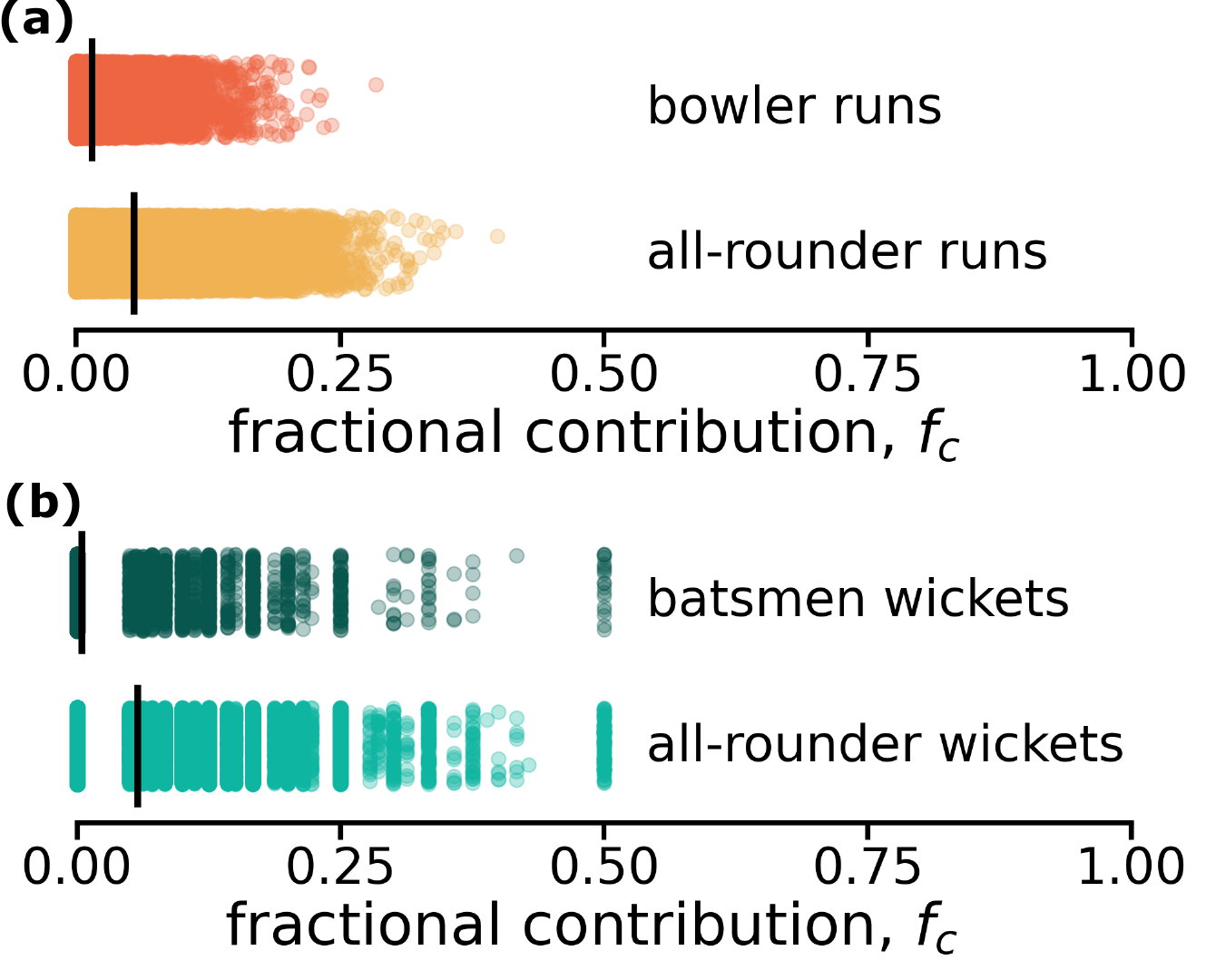}
    \caption{Cross-category contribution of various types of players in batting and bowling. (a) A bowler contributes 1\% of team runs, while an all-rounder contributes 5\% of team runs on average. A batsman contributes 7\% of team runs as seen in the main text. (b) A batsman contributes to 0.3\% of team wickets, while an all-rounder contributes to 5\% of team wickets. A designated bowler contributes 10\% of team wickets as seen in the main text.}
    \label{fig:cross_category}
\end{figure}

\newpage
\section{Individual contributions to team success}
\begin{figure}[!h]
    \centering
    \includegraphics[width=0.7\columnwidth]{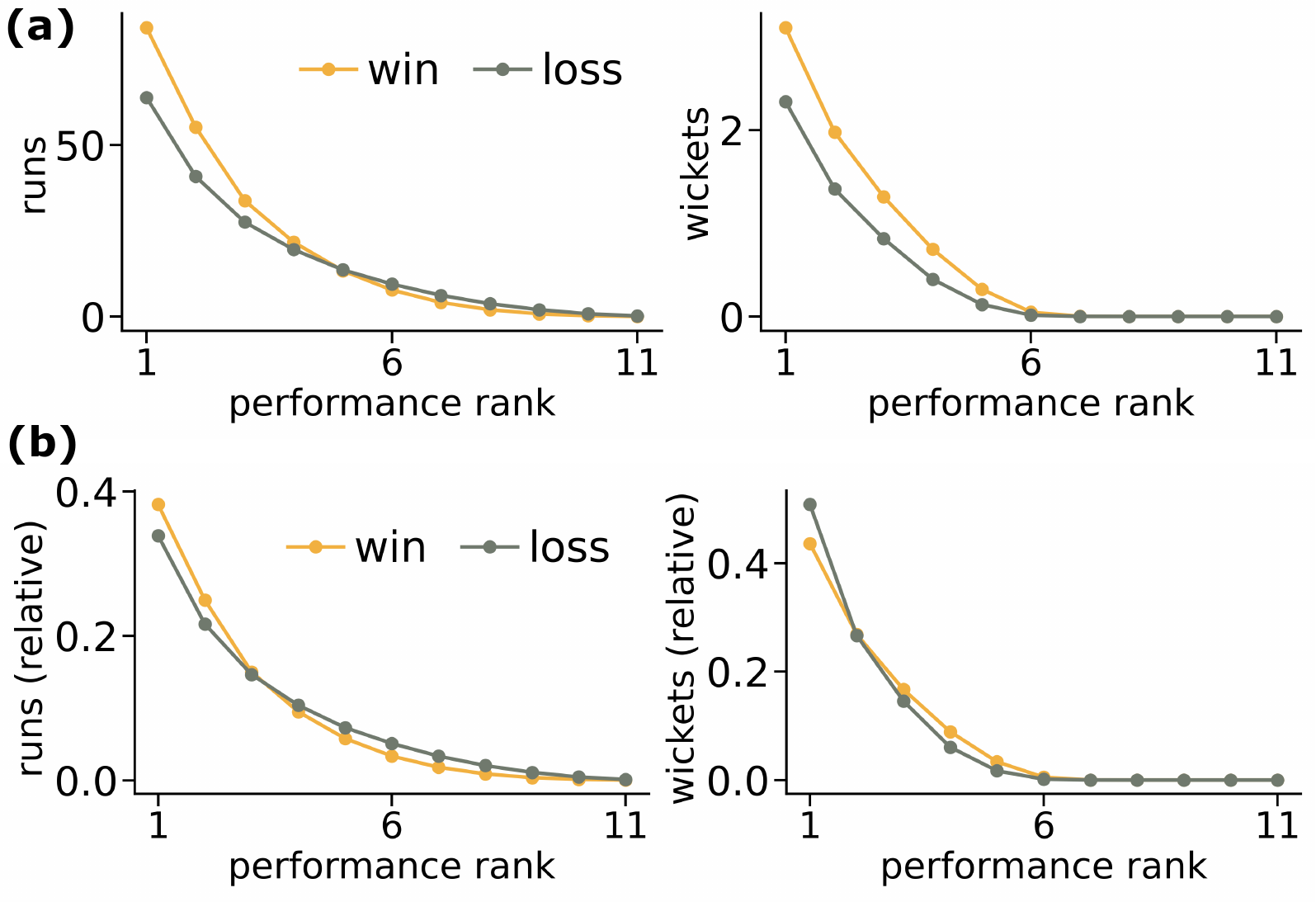}
    \caption{Individual contributions to the team in absolute (top) and relative scales (bottom) for winning and losing teams. We notice that the top performers (in absolute scale) for the winning team always outperform the top performers of the losing team. On the other hand, when we look at the relative contributions, top batsmen from the winning team contribute higher than their losing counterparts. However, the top bowler from the losing team contributes more to his team than that of the top bowler from the winning team.}
    \label{fig:individual_to_team}
\end{figure}

\end{document}